\documentclass[10pt,conference]{IEEEtran}

\usepackage[hang,flushmargin,stable]{footmisc}

\usepackage{graphicx}
\usepackage{balance}  %
\usepackage{graphics} %
\usepackage{times}    %
\usepackage{url}      %
\usepackage{paralist} %

\usepackage[numbers,sort&compress]{natbib}

\usepackage{amssymb}
\usepackage{amsmath}
\usepackage{colortbl}
\usepackage{color}
\usepackage[font=bf]{caption}
\usepackage{url}
\usepackage{paralist}

\definecolor{darkgreen}{RGB}{0,0,150}

\usepackage{url}
\usepackage{color}
\definecolor{darkblue}{RGB}{0,0,120}
\usepackage[bookmarks=false, colorlinks=true, plainpages=false,  linkcolor=darkblue,   citecolor=darkblue, urlcolor=darkblue, filecolor=darkblue]{hyperref}
\usepackage{breakurl}

\newcommand{\descr}[1]{\vspace{0.3cm} \noindent \textbf{#1}}
\newcommand{\fake}[1]{\vspace{0.25cm} \noindent \emph{#1}}

\usepackage[lofdepth,lotdepth]{subfig}

\begin{document}

\title{An Exploratory Ethnographic Study of Issues and Concerns with Whole Genome Sequencing$^*$\thanks{$^*$A preliminary version of this paper appears in USEC 2014.}}

\author{\IEEEauthorblockN{Emiliano De Cristofaro}
\IEEEauthorblockA{University College London$^\dag$\\
me@emilianodc.com}}

\IEEEoverridecommandlockouts

\maketitle

\renewcommand{\thefootnote}{\fnsymbol{footnote}}
\footnotetext{$^\dag$Work done in part while at PARC (a Xerox Company).}
\renewcommand{\thefootnote}{\arabic{footnote}}

\begin{abstract}
Progress in Whole Genome Sequencing (WGS) will soon allow a large number of individuals to have their genome fully sequenced. This lays the foundations to improve modern healthcare, enabling a new era of personalized medicine where diagnosis and treatment is tailored to the patient's genetic makeup. It also allows individuals motivated by personal curiosity to have access to their genetic information, and use it, e.g., to trace their ancestry. 
However, the very same progress also amplifies a number of ethical and privacy concerns, that stem from the unprecedented sensitivity of genomic information and that are not well studied.

This paper presents an exploratory ethnographic study of users' perception of privacy and ethical issues with WGS, as well as their attitude toward different WGS programs. We report on a series of semi-structured interviews, involving 16 participants, and analyze the results both quantitatively and qualitatively. Our analysis shows that users exhibit common trust concerns and fear of discrimination, and demand to retain strict control over their genetic information. Finally, we highlight the need for further research in the area and follow-up studies that build on our initial findings.
\end{abstract}

\section{Introduction}
In the past half a century, research on DNA sequencing -- the process of determining the precise order of nucleotides within a DNA molecule -- has made tremendous progress. 
The first complete human genome was sequenced in 2003 within a 13-year, \$3B project known as the   Human Genome Project~\cite{levy07,wheeler08}.
Thereafter, there has been an exciting race toward faster, cheaper, and more accurate
Whole Genome Sequencing (WGS) technologies.
Costs have quickly plunged to 
\$250K already in 2008~\cite{quail2008large} %
and to a few thousand dollars in the 2010's~\cite{illumina4k,drmanac2010human}.
The long-anticipated \$1,000 threshold was breached, in January 2014, by the San Diego-based company Illumina~\cite{illumina14}.
The landscape of companies and technologies in WGS is
fast-evolving, nonetheless, it is not far-fetched to predict that, in 5-10 years, most individuals in developed countries will have access to WGS.

The emergence of affordable WGS technologies represents an exceptional breakthrough,
due to the expected medical and societal implications.  Biomedical experts anticipate that advances in WGS will unlock the full potential of {\em personalized medicine}~\cite{ginsburg2009}, i.e., the practice of tailoring
pre-symptomatic examinations, diagnosis, and treatment to patients' genetic features.
Arguably, the availability of patient's fully sequenced genome enables clinicians, doctors, and testing facilities to run a number of complex genetic tests in a matter of seconds, using specialized computational algorithms (as opposed to more expensive and slower laboratory tests). However, 
due to  the extreme sensitivity of DNA data~\cite{collins2001implications},
WGS also comes at the risk of amplifying important security, privacy, and ethical concerns.
These issues have been recently under the scrutiny of several researchers~\cite{ayday2013chills} and policy experts~\cite{president} -- we overview them in Section~\ref{subsec:privacy}.

\descr{Motivation.} The emergence of personal genomic tests as well as the anticipated
availability of affordable WGS technologies, along with the associated privacy and ethical
concerns, motivate the need for better understanding 
users' perceptions and attitudes. For instance, it is desirable to assess whether or not
privacy perceptions and concerns experienced by patients
correspond to what the scientific community would expect,
as well as how to identify effective mechanisms to communicate 
potential privacy risks associated with genomic information and its disclosure.
Even more importantly, much of the anticipated progress in personalized medicine, and human genome research in general, will not happen without experimental research relying on large cohorts of volunteers willing to share their genetic 
material.\footnote{E.g., the US/Canada Personal Genome Project~\cite{PGP} and 
the UK 100K Genome initiative~\cite{cameron}  rely on volunteers that agree to have their genomic data  made publicly available.}
Therefore, it is fundamental to understand---and address---individuals' concerns and fears (such as embarrassment or discrimination) in order to retain their trust and their willingness to participate.

A few studies (reviewed Section~\ref{sec:related}) have analyzed individuals' response to learning the results of some genetic tests and potential discrimination concerns associated with them, but there is very little understanding of these concerns in the context of WGS, and the new scenarios enabled by it.

\descr{Contributions.} This paper presents the results of a series of semi-structured interviews, involving 16 participants. We aim to assess the perception of privacy and ethical issues with WGS as well as the attitude toward different sequencing programs.
Results are analyzed both quantitatively and qualitatively, yielding a few interesting findings related to the issues of control and trust,
as users exhibit common fear of discrimination and demand to strictly control how their genetic information is used.
Our exploratory study is vital to guide follow-up (possibly larger) ethnographic studies on the topic, which we plan to conduct as part of future work.

\descr{Paper Organization.} The rest of the paper is organized as follows:
next section presents background information about genomics and 
privacy/ethical concerns, then, Section~\ref{sec:methodology} presents our study design and participants' demographics, while Section~\ref{sec:results} analyzes
study results, both quantitatively and qualitatively. 
After reviewing related work in Section~\ref{sec:related},
the paper concludes in Section~\ref{sec:conclusion}.

\section{Background}\label{sec:background}
This section presents some background information on genomics as well as privacy
threats stemming from advances in Whole Genome Sequencing (WGS).

\subsection{Genomics Primer}\label{subsec:genomics}

We start with discussing Personalized Medicine -- the practice of tailoring
pre-symptomatic examinations, diagnosis, and treatment to individuals' genetic features.
Genomic tests today are increasingly and more effectively used in healthcare.
For instance,  testing for the {\em tpmt} gene is already required prior to prescribing some drugs used for treating childhood leukemia~\cite{abbott2003}. 
Similarly, doctors prescribing Zelboraf (Roche's treatment for skin cancer) need to test the patient for the BRAFV 600E mutation. %
Other analogous examples include testing for mutations in the Philadelphia chromosome (in Acute Lymphoblastic Leukemia patients) or in BRCA1/BRCA2 genes (in breast and ovarian cancer patients). 

Also, a few commercial companies, such as, 23andMe.com, provide  customers with reports on predisposition to diseases and conditions, even though they do not yet rely on WGS. (Specifically, 23andMe.com provides customers with a low-cost (about \$100) report assessing their genetic risk toward a number of diseases and conditions, by testing for almost a million specific Single-Nucleotide Polymorphisms (SNPs)~\cite{dbSNP,stenson}.)
Also, several drugs (e.g., for cancer, HIV, or thrombosis treatment) are paired today with genetic tests needed to assess either the correct dosage or their expected effectiveness~\cite{abbott2003,900,oldenburg2007vkorc1,trastuzumab}. 

Experts estimate that about a third of the 900 cancer drugs currently in clinical trials will come to the market with a DNA or other molecular test attached~\cite{900}. Although most predominant, cancer treatment is only
one of the application fields of personalized medicine. For instance, for some cardiac patients, recovery from a common heart procedure can be complicated by a single gene responsible for drug processing, and selection of blood thinner drugs should depend on whether or not patient holds such a gene mutation~\cite{heart}. 

In general, advances in WGS facilitate the understanding of the impact of genetic variations on the response to medications. For instance, 
researchers have  discovered genes  encoding Cytochrome P450 enzymes, which metabolize neuroleptic medications to improve drug response and reduce side-effects~\cite{cichon2001pharmacogenetics}, and also genes
involved in the action and metabolism of warfarin (coumadin) -- a medication used as an anticoagulant~\cite{oldenburg2007vkorc1}.

\medskip The availability of whole human genomes will also facilitate a number of genetic tests that today are performed in vitro, by reducing costs and time.
For instance, \emph{ancestry and genealogical tests} allow individuals to trace their lineage by
analyzing their genomic information (the scope of such tests being often quite heterogeneous). 
Ancestry testing
is increasingly popular, e.g., to map one's own 
genetic heritage and/or find known ancestry. Several commercial entities (e.g., 23andMe.com and Ancestry.com) 
provide customers with reports on their genetic ancestry. They maintain a 
collection of sample genomes from individuals belonging to different ethnic groups, and compare them against
their customers' genomic information to understand how they relate to known ethnic groups.

\emph{Genetic compatibility tests} are used to let (potential or actual) partners assess the possibility of
transmitting to their children a genetic disease with Mendelian inheritance~\cite{mendelian}.
For instance, {\em Beta-Thalassemia minor} causes red cells to  be smaller than average, due to a mutation in the {\em hbb} gene. It is called {\em minor} when
the mutation occurs only in one chromosome, while
the {\em major} variant---that occurs when both chromosomes carry the mutation---is
likely to result in premature death. Therefore, if both partners carry the {\em minor} form, there is a non-negligible chance that their child will carry the {\em major} variant.

\begin{table*}[thb]
\centering
\small
{\parbox{1\textwidth}{%
\begin{minipage}[t]{0.2\linewidth}\centering
\begin{tabular}{| l | c | r |}
\hline
{\bf Age} & {\bf N} & {\bf \%} \\
\hline
18--24 & 2 & 12.5\%\\
25--34 & 7 & 43.7\%\\
35--44 & 3 & 18.7\%\\
45--54 & 1 & 6.2\%\\
55--64 & 1 & 6.2\%\\
65--75 & 2 & 12.5\%\\
\hline
\end{tabular} 
\label{age-int}
\end{minipage}%
\begin{minipage}[t]{0.25\linewidth}\centering
\vspace{-1.3cm}
\begin{tabular}{| l | c | r |}
\hline
{\bf Degree} & {\bf N} & {\bf \%}\\ \hline
College	& 4 & 25\%\\
Master & 8 & 50\%\\
PhD & 4 & 25\%\\
\hline
\end{tabular}
\label{edu-int}
\end{minipage}
\begin{minipage}[t]{0.25\linewidth}\centering
\vspace{-1.3cm}
\begin{tabular}{| l | c | r |}
\hline
{\bf Yearly Income} & {\bf N} & {\bf \%}\\ \hline
Less than \$50K	& 3 & 18.7\%\\
\$50K-\$75K & 3 & 18.7\%\\
More than \$75K & 10 & 62.5\%\\
\hline
\end{tabular}
\label{inc-int}
\end{minipage}
\begin{minipage}[t]{0.3\linewidth}\centering
\vspace{-1.3cm}
\begin{tabular}{| l | c | r |}
\hline
{\bf Westin  Index} & {\bf N} & {\bf \%}\\ \hline
Unconcerned	& 4 & 25.0\%\\
Pragmatist & 7 & 43.7\%\\
Fundamentalist & 5 & 31.2\%\\
\hline
\end{tabular}\\
\label{wpi-int}
\end{minipage}%
\caption{Participants breakdown by age, education, (personal) yearly income, and Westin's Privacy Index~
\cite{westin}.}
\label{tab:demor}
}}
\end{table*}

\subsection{WGS Privacy and Ethical Threats}\label{subsec:privacy}
While advances in genomics and sequencing technologies
promise to unlock a wide range of medical and societal benefits,
they also risk to amplify security, privacy, and ethical concerns~\cite{ayday2013chills}.

The human genome not only uniquely and irrevocably identifies its owner, but also contains information about 
ethnic heritage, predisposition to numerous diseases and conditions, including mental 
disorders~\cite{correlated,canli2007emergence,fumagalli2009parasites}.
Furthermore, due to its hereditary nature, disclosing one's human genome also implies,
to a certain extent, disclosing the genomes of close relatives~\cite{humbert2013addressing}.

Traditional approaches to privacy, such as de-identification or aggregation, might not be effective when applied to genomic information. 
\cite{gymrek2013identifying} demonstrates the feasibility of re-identifying DNA 
donors from a public research database using information from popular genealogy websites. Additional results on DNA and genome re-identification include~\cite{malin2006,malin2001re,malin2004,homer2008,wang2009learning,Gitschier_2009}.

As the human genome contains detailed information about susceptibility to 
somatic and mental conditions, ethnicity, etc., its disclosure is often
associated to fears of {\em eugenism}, i.e., genetic discrimination, thus potentially
affecting social dynamics as well as hiring and healthcare practices. (See~\cite{geller1996individual} for a survey of self-reported genetic discrimination.)

These concerns creates the need for \emph{informed consent} to guard against surreptitious DNA testing by requiring authorities and companies to obtain written permission from citizens before collecting, analyzing, storing or sharing their genetic information (e.g., preventing people
from collecting hair or saliva samples and maliciously sequencing the victims' genome). However, on the other hand,
such privacy-restrictive measures are often regarded as a potential obstacle to genomic research. Scientists typically sequence DNA from thousands of people to discover genes associated with particular diseases, thus, the informed consent restriction would mean that large genomic datasets could not be re-used to study a different disease -- researchers would either need to destroy the data after each study, or track down thousands of former subjects for new authorizations~\cite{fear}.

As mentioned earlier, much of the anticipated progress in personalized medicine, and human genome research in general, will not happen without experimental research relying on large cohorts of volunteers willing to share their genetic 
material. Therefore, it is fundamental to analyze---and address---individuals' concerns and fears in order to retain their trust and their willingness to participate.

\section{Interview Methodology}\label{sec:methodology}

Our study consisted of a series of semi-structured interviews with 16 participants (8 female, 8 male), conducted in Summer 2013.
It was reviewed and approved by PARC's Institutional Review Board (IRB).
Participants were recruited using social networks and local mailing lists, announcing
a study on  the ``knowledge and perception of DNA testing.''

\subsection{Participants}
The 16 participants were US-based and ranged in ages from 18 to 74.
We collected information about participants' personal yearly income, and assessed their Westin's Privacy Index~\cite{westin}, %
which classifies users among privacy fundamentalists, pragmatists, and unconcerned, according to their responses to a simple three-question survey.
Demographics breakdown is reflected in Table~\ref{tab:demor}.
Study participants volunteered to participate in the study, i.e., they received no monetary incentive: we decided to do so in order to recruit motivated users (who were also likely to possess some understanding of genetics).

\descr{NOTE:} We deliberately decided to recruit users with college degrees
and more participants in the 25-34 age range as they constitute the representative population of early adopters of personal genomic tests and WGS, as suggested by~\cite{facio2011motivators,npr}. This is crucial for our study since WGS is still at an early (research) phase and is not yet available to ``regular'' users.

\subsection{Experiments}
Interviews lasted approximately 30 minutes (on average) and consisted of  three parts: (1) first, we asked 
participants to provide 1-2 examples of genetic tests and describe their familiarity with genetics; (2)
we interviewed participants while guiding them through a set of slides
depicting a few hypothetical scenarios (this constitutes the core of our study, see details below);
(3) in order to collect
demographic information and assess their Westin Privacy Index~\cite{westin},
participants were asked to fill out a short survey.

We aimed to conduct the four experiments presented in Table~\ref{tab:summary} below.

\begin{table}[h!]
\centering
\small
\vspace{0.2cm}
    \begin{tabular}{|c|l|}
	\hline
	{\bf Exp. A}  & Assessing the perception of today's \\
							& genetic tests\\
	\hline
	{\bf Exp. B}  & Comparing the attitude toward different\\
							& WGS programs\\
	\hline
	{\bf Exp. C}  & Assessing the perception of potential\\
							& privacy/ethical issues with WGS\\
	\hline
	{\bf Exp. D}  & Comparing the response to medical, \\
				 & genomic, and personal information loss\\
	\hline
	\end{tabular}
	\caption{Overview of study experiments and goals.}
	\label{tab:summary}
\end{table}

\subsubsection{Experiment A: Perception of genetic tests}

 Participants were shown the following six examples of genetic tests. 
Note that the order in which the test scenarios were actually presented was randomized
across multiple participants.

\fake{(A.1) Disease Predisposition (Doctor.)} Scott goes to see his primary care doctor. The doctor asks Scott to run some genetic
tests to assess (genetic) predisposition to certain diseases. After collecting a DNA sample
(e.g., saliva), the clinic runs some tests on Scott's DNA, and the doctor tells Scott that he is predisposed
to type-2 diabetes, but not to Alzheimer's.

\fake{(A.2) Genetic Compatibility.} Emma and Scott are planning to have kids. Emma has Beta-Thalassemia {\em Minor}, a genetic disorder
inherited by only one of her parents, which causes her red cells to be smaller than usual but poses no 
critical health threat to her. However, if Scott also has the same disorder, there is a chance that
their kids will have the {\em Major} form, which may cause premature death. Therefore,
Scott is advised to take a genetic test to make sure that he doesn't have the disorder.

\fake{(A.3) Genetic Ethnicity.} Scott is a customer of a genealogy company, such as, Ancestry.com, which 
analyzes genetic ethnicity and reveals where one's ancestors were form. Scott needs
to send a DNA sample (e.g., saliva) to the company, which, after running some tests,
tells Scott that his ethnic origins are 39\% Scandinavian, 17\% Central European, etc.

\fake{(A.4) Disease Predisposition (Company.)} Scott is a customer of a personal genomics company, such as, 23andMe.com, which
provides its customers with a detailed report on their genetic chances
of getting a number of diseases, like type-2 diabetes, Alzheimer's, schizophrenia, etc.
For each disease, the report includes a risk percentage, the known average risk percentage,
and a confidence score. Scott needs to send a DNA sample (e.g., saliva) to the company,
which, after running some tests, provides Scott access to a Web interface with the report.

\fake{(A.5) Correct Drug Dosage.} Scott needs to take a blood thinning drug like warfarin. His doctor, in order to better assess the right dosage for him, requests that Scott takes a genetic test. After collecting a DNA sample
(e.g., saliva), the clinic runs some tests and, based on the results, the doctor prescribes the
exact warfarin dosage for Scott.

\fake{(A.6) Correct Cancer Treatment.} 
Scott is diagnosed with cancer. His doctor, prior to propose a treatment plan, requests
that Scott takes a genetic test to assess whether or not he carries a certain DNA mutation.
After collecting a DNA sample (e.g., saliva), the clinic runs some tests and, based on the results, 
the doctor suggests what the best treatment for Scott is.

\smallskip
\noindent 
After showing the above test scenarios, we presented a recap slide summarizing them.
We then asked interviewees to compare those scenarios; specifically, we raised
questions like {\em ``If you were in Scott's shoes, would you do all of these tests? Which ones would you not?''}
as well as more specific ones like {\em ``If you were Scott, which tests would you feel more inclined to do? Which tests
more reluctant to?''} Finally, we asked participants to actually {\bf\em rank the six tests} from the one they felt most inclined to,
to the one they felt most reluctant.

\subsubsection{Experiment B: Attitude toward WGS}
First, we presented to the participants one slide describing (with both text and pictures) Whole Genome Sequencing (WGS), i.e., the process of digitizing the complete DNA sequence of a human genome.
We told participants that once a whole human genome is sequenced, all genetic tests, including those presented in Experiment A., could be performed on the sequenced data, using computer techniques. We also told them that computer techniques could be faster and cheaper than collecting a sample (and using expensive lab equipment) each time one needs a genetic test.

Afterwards, we presented (again, in random order) the following three scenarios.

\fake{(B.1) WGS with Healthcare Provider.} Emma goes to see her primary care doctor, who tells her about
the provider's experimental Whole Genome Sequencing program. He explains to her that, by having her genome
sequenced, Emma may learn her genetic predisposition to diseases and conditions. He also tells her
that the provider will keep her sequenced genome along with her medical information, so that they can 
use it to run genetic tests when needed for medical reasons. Also, the provider will offer a discount
on the next healthcare bill.

\fake{(B.2) WGS with Personal Genomics Company.} Emma decides to have her genome sequenced
by using a personal genomics company, such as, 23andMe.com.
The company charges her \$100 to sequence her genome, give her a report about predisposition to diseases
and conditions, and tell her about her genetic ancestry. Also, the company offers to store the data for her:
if she wants to authorize her doctor to run some tests, she can grant him permission to contact the company.

\fake{(B.3) Data-only WGS (DVD).} Emma decides to have her genome sequenced
by using a sequencing laboratory, which charges her \$100 to sequence her genome and send her
a DVD with the data.

\smallskip
\noindent 
After being presented with the above scenarios, participants were shown a recap slide.
We then invited them to compare these scenarios; specifically, we 
asked {\em ``If you were Emma, would you participate in any of the three programs?} and {\em ``Which program(s) would you feel more inclined to participate in? Which one(s) less inclined?}
Finally, we asked participants to {\bf rank} the three scenarios from the one they felt most inclined to,
to the one they felt most reluctant to.

\begin{table*}[thb]
\centering
\footnotesize
\subfloat[][Ranking average and standard deviation for Exp A.]{
\begin{minipage}[b]{0.48\linewidth}\centering
\small
    \begin{tabular}{|l|c|c|}
	\hline
    \textbf{Genetic Tests: More to less inclined \hfill} & \textbf{Avg} & \textbf{Stdev} \\
	\hline
    (A.6) Correct Cancer Treatment & 5.81  & 0.39 \\
    (A.5) Correct Drug Dosage & 4.63  & 0.70 \\
    (A.2) Genetic Compatibility & 4.06  & 1.25 \\
    (A.1) Disease Predisp. (Doctor) & 2.63  & 0.99 \\
    (A.4) Disease Predisp. (Company) & 2.13  & 0.70 \\
    (A.3) Ancestry Testing & 1.75  & 1.09 \\
	\hline
	\end{tabular}
  	\label{tab:expA}
\end{minipage}
}
\hfill
\subfloat[][Ranking average and standard deviation for Exp B.]{
\small
\begin{minipage}[b]{0.48\linewidth}\centering
    \begin{tabular}{|l|c|c|}
	\hline
    \textbf{WGS Programs: More to less inclined $\;\;$ } & \textbf{Avg} & \textbf{Stdev} \\
	\hline
	(B.3) Data-only (DVD)    & 2.68  & 0.58 \\
    (B.1) Healthcare Provider     & 2.00  & 0.71 \\
    (B.2) Personal Genomics Company    & 1.31  & 0.46 \\
    \hline
    \end{tabular}
  \label{tab:expB}
\end{minipage}\vspace{0.6cm}
}\\
\subfloat[][Ranking average and standard deviation for Exp C.]{
\begin{minipage}[b]{0.48\linewidth}\centering
\small
    \begin{tabular}{|l|c|c|}
	\hline
    \textbf{Incidents: More to less discomfort} & \textbf{Avg} & \textbf{Stdev} \\
	\hline
	(C.1) Labor Discrimination & 3.31  & 0.58 \\
    (C.2) Health Insurance Discrimination & 3.00  & 0.94 \\
    (C.3) Sequenced Genome Hacked & 2.56  & 0.93 \\
    (C.4) Sibling Donating Genome to Science & 1.13  & 0.33 \\
    \hline
    \end{tabular}
  \label{tab:expC}
\end{minipage}
}
\hfill
\subfloat[][Ranking average and standard deviation for Exp D.]{
\begin{minipage}[b]{0.48\linewidth}\centering
\small
    \begin{tabular}{|l|c|c|}
	\hline
    \textbf{Information loss: More to less frightened} & \textbf{Avg} & \textbf{Stdev} \\
	\hline
    (D.1) Identity Theft & 3.50  & 0.63 \\
    (D.3) Emails and Pictures & 2.63  & 1.61 \\
    (D.4) Sequenced Genome & 2.00  & 0.63 \\
    (D.2) Medical Records & 1.88  & 0.48 \\
    \hline
    \end{tabular}
  \label{tab:expD}
  \end{minipage}
}
  \caption{Summary of Experiments.}
\end{table*}

\subsubsection{Experiment C: Perception of privacy/ethics issues with WGS}

 In this experiment, we presented (in random order) the following 
four hypothetical cases.

\fake{(C.1) Labor Discrimination.} Emma is interviewing for a job and her potential employer
requests to run a drug test and to access her genome. The employer finds out that Emma
has predisposition to cancer and decides not to hire her.

\fake{(C.2) Health Insurance Discrimination.} Emma wants to join a health insurance plan.
The provider requests access to her medical record as well as her genome, and finds out
that she has predisposition to leukemia. As a result, Emma needs to pay a higher premium.

\fake{(C.3) Sequenced Genome Stolen by Hacker.} A hacker steals the file with Emma's sequenced
genome and finds out her ethnicity/ancestry and that she has genetic predisposition to Huntington's disease.

\fake{(C.4) Sibling Donating Genome to Science.} Emma's brother, Jack, decides to donate his genome to science (for medical research purposes).
Since Emma's and Jack's genomes are 99.9\% identical, Jack is actually ``donating'' Emma's genome as well
but has not asked for her permission.

\smallskip
\noindent 
After being presented with the above scenarios, participants were shown a recap slide.
We then asked to compare them; specifically, we 
asked: {\em ``If you were Emma, what incident(s) would give you the most discomfort? Which one(s) the least?}
Finally, we asked participants to {\bf\em rank the four scenarios} from the one giving them the most discomfort,
to the least.

\subsubsection{Experiment D: Response to information loss}
In the last experiment, participants were asked to {\bf\em rank the following four 
incidents} (presented in random order) from the one they found most frightening to the least.

\fake{(D.1) Identity.} A hacker steals your financial information and your social security number,
i.e., so-called ``identity theft.''

\fake{(D.2) Medical records.} A hacker hacks into your clinic's databases and steals all your medical records.

\fake{(D.3) Emails and Pictures.} A hacker hacks into your computer and steals your personal emails and pictures.

\fake{(D.4) Sequenced Genome.} A hacker steals the file containing your sequenced genome.

\section{Study Results}\label{sec:results}
We now present the results of our study, via a quantitative and qualitative analysis.

\subsection{Theme 1: Trust} 

During Experiment A, we asked participants to rank six genetic tests
from the one they felt most inclined to, to the least.
We report the average rankings, along with standard deviation, in Table~\ref{tab:expA}. (Rankings are on a 1-6 scale, with 6 being the
case participants felt most inclined to, 1 the least.)

(A.6), i.e., correct cancer treatment, emerged as the test to which participants felt most inclined, with average
ranking 5.81 (out of 6). 13/16 participants 
ranked (A.6) as the top one, and all of them among the top two. 
Next follow (A.5), i.e., correct drug dosage, and (A.2), i.e., genetic compatibility, with, respectively, 15/16 and 13/16 of participants ranking it among the top three. 

At the other extreme, (A.3), ancestry testing, emerged as the test scenario most of participants (10/16) felt the least
inclined to, with an overwhelming majority (12/16) ranking it among the bottom two. 
However, only 2 participants mentioned that they would not want to know their ancestry, while 8 reported that they would not mind discovering it but found medical tests much more relevant.
Average ranking for (A.3) was 1.75, with a relatively high standard deviation of 1.09.

(A.1), i.e., disease predisposition (doctor), and (A.4), i.e., disease predisposition (company),
were ranked relatively similar, mostly in the middle of the ranking scale.
The former obtained a slightly higher average, i.e., 2.63 vs 2.13, although
their difference was not statistically significant.

\smallskip
Participants exhibited a preference toward tests they considered immediately
beneficial for their health (or that of their kids), such as, (A.6), (A.5), and (A.2).
In fact, the rankings for (A.2), the test with lowest average in this category, 
and those for (A.1), the test with highest average outside this category,
are statistically significant, as per Mann-Whitney U Test (U $=210.5$, $n_1=n_2=16$, $P<0.01$, two-tailed).

Finally, it is interesting to observe the difference between (A.1) and (A.4): recall that, in both cases, the patient discovers the genetic predisposition to certain diseases. There were 9/16 participants ranking (A.1) higher than (A.4), and 5 out of 9 justified this choice due to a lower {\em trust} in a commercial company. For instance, P3 mentioned that she {\em ``would never trust a website with my genome.''} The other 4 participants instead mentioned that they would prefer a doctor to explain the test results, as the effects of genetic disease predisposition were not completely clear to them.
On the other hand, 7/16 participants ranked (A.4) higher than (A.1): 2 of them justified this as they do not like to go to the doctor; 2 of them preferred a broader analysis of all predispositions; and, the remaining 3 did not want their healthcare provider to know about their predisposition, owing to fears of coverage denial 
or higher premiums.

\subsection{Theme 2: Control}
In Experiment B, we aimed to understand and analyze participants' preference among
three different options for Whole Genome Sequencing (WGS) programs.
To this end, we asked participants to rank these options from the one they felt most
incline to, to the least. The average rankings, along with standard deviation, are reported in 
Table~\ref{tab:expB}. (Rankings are on a 1-3 scale, with 3 indicating the most favorite option, 1 the least favorite.)

The favorite option for a hypothetical WGS program was (B.3), i.e., data-only WGS (DVD),
with more than two thirds of the participants (12/16) ranking it at very the top (average ranking 2.68 out of 3).
Whereas, half of the participants (8/16) ranked option (B.1), WGS with healthcare provider, as their second favorite.
Clearly, option (B.2), WGS with personal genomics company, seems to be the least preferred (1.31 average ranking), with two thirds of the participants (11/16)
ranking it at the bottom, and 16/16 ranking it among the bottom two.
In fact, the difference between the rankings of (B.2) and those of (B.1) is significant as per
Mann-Whitney U Test (U $=194, n_1=n_2=16, P<0.05$, two-tailed).

\smallskip
We noticed that several participants, 12/16, mentioned that 
they {\em wanted to feel in control} of their genomic data.
In particular, {\em ``control''} was often used to justify the choice of ranking (B.3) as the preferred WGS option. Recall that
12/16 participants expressed preference toward
(B.3), i.e., obtaining their sequenced genome on a DVD via a testing facility that would not retain data.
Among these, 7 participants (6 male, 1 female) mentioned control as their {\em main} motivation for doing so, whereas, 5 users (all female)
gave other reasons.\footnote{This also suggests a correlation between gender
and preferring (B.3) because of increased control, with $\chi^2(1, N=32)=8.57 ~ p<0.01$.}

P10 said {\em ``I would feel in control of my data,''}, while he would not trust it with the healthcare provider, as
{\em ``they could hold it against me.''} And (B.2) would be {\em ``even worse as the company would not be subject to the same standards as a healthcare provider.} Similarly, P14 reported she {\em ``would like to own and control my genome, or at least let my family do''} and ranked (B.1) at the bottom, mentioning a complete lack of control in giving the information to a healthcare provider.
Also, P12, P15, and P16 raised control as the main motivation to prefer (B.3); nevertheless, P12 reported that he {\em ``wouldn't mind (B.1) if they can guarantee the same level of control,''} while both P15 and P16 mentioned that it would be necessary to trust the sequencing facility not to retain the raw data. 

Next, we noticed that 7/16 participants responded negatively to the idea of receiving a discount on their next health bill in exchange for 
letting their provider sequence their genome and give a list of disease predispositions. P6 reported that {\em ``the discount looks fishy [...] are they going to make a deal with my genome?''} P9 said {\em ``I don't care if I get a discount, I would prefer that they promise not to deny me coverage.''}
On the other hand, P5, who ranked (B.3) higher than (B.1), pointed out that she would consider changing her mind if the discount was significant ({\em ``higher than \$1,000.''})

Additional noteworthy remarks include P4 and P8 worrying of losing the DVD with the sequenced genome and P16's fear of being served personalized ads by a personal genomics company performing the sequencing.

\subsection{Theme 3: Discrimination}
Experiment C aimed to assess perception of potential privacy and ethical issues associated
with genomic tests and availability of whole genomes.
Participants were asked to compare four different incidents and rank them
from the one giving them the most  discomfort, to the least. 

The average rankings, along with standard deviation, are reflected in 
Table~\ref{tab:expC}. (Rankings are on a 1-4 scale, with
4 indicating the most frightening case, and 1 the least frightening.)
{\em Note: unlike Experiment A and Experiment B,
a higher ranking in Experiment C corresponds to a negative feeling.}

(C.1), i.e., labor discrimination, and (C.2), i.e., health insurance discrimination,
represented the incidents giving most discomfort to 
the participants, with 12/16 participants ranking either one as the top.
More specifically, almost one third of the participants (10/16)
ranked (C.1) and (C.2) as the two incidents giving them most discomfort,
owing to related discrimination issues.
Their average rankings were both above 3.00 (out of 4).
(C.1) actually seems to be the most discomforting scenario as 15/16 participants placed
it in the top two, compared to 11/16 for (C.2).

Then, 4/16 participants felt that (C.3), i.e., sequenced genome hacked, was the most discomforting scenario. 2 of these 4 users actually reported that (C.3) would imply (C.1) and (C.2), as a hacker could publish the genomic information or sell it to employers and healthcare providers. Whereas, the other 2 participants felt that they should be protected by the law as for (C.1) and (C.2).

(C.4), i.e., sibling donating genome to science, represented the case yielding the least 
discomfort, with 14/16 participants ranking it at the bottom, and 16/16 among the bottom two. 
The difference between rankings of (C.4) and those of (C.3), the second least discomforting case,
was statistically significant, as per the Mann-Whitney U Test (U $=238.0$, $n_1=n_2=16$, $P<0.001$, two-tailed).

\smallskip
Concerns related to discrimination were raised by several participants when motivating the rankings in Experiment C, but also, indirectly,
when discussing their concerns with respect to Whole Genome Sequencing.

10/16 participants ranked (C.1) higher than (C.2): when asked why, 5 out of these 10 participants reported that somehow (C.2) was not too surprising. P5,  P6, and P16 actually said that healthcare cost discrimination already happens today, so (C.2) would be somewhat expected. P16 also mentioned that (C.2) {\em ``is understandable, they will be the ones to pay, and they already take into account predisposition and pre-existing conditions.''}

Discomfort from (C.1) was often paired with an unfairness feeling. P7 pointed out that predisposition {\em ``does not necessarily mean that I will get the disease,''} and P1 that predisposition {\em ``doesn't mean that I won't be able to perform my job.''}

Finally, (C.4) was the least frightening case for 14/16 participants: 9 of these said they were not concerned since the genome is donated to science, for a good cause, and 5 trusted the research lab having access to their genome, since, as pointed out by P13, {\em ``labs are subject to ethical reviews and will not share data with anyone.''}

\subsection{Theme 4: Damage from Information Loss}

The last experiment aimed to compare participants' response to incidents
involving information loss. We asked them to rank four hypothetical cases
from the one they considered most frightening to the one they considered
least. The average rankings, along with standard deviation, are reported in 
Table~\ref{tab:expD}. Rankings are on a 1-4 scale, with
4 indicating the most frightening case, and 1 the least frightening.) Again, unlike Experiment A and Experiment B,
and similar to Experiment C, a higher ranking in Experiment D corresponds to a negative feeling.

(D.1) represented the most frightening case for most participants,
with an average ranking of 3.50 (out of 4).
Indeed, 10/16 participants ranked it as the most frightening scenario 
and 15/16 among the top two. 
Then follows (D.3), with 6/10 participants ranking it at the top.
Observe that the average ranking is relatively low (2.63) with a high standard deviation (1.61);
almost half of the participants (7/16) ranked it among the bottom two cases, and 5/16 at the very bottom.

(D.4) and (D.2) were ranked similar, with average ranking, resp., 2.00 and 1.88, and no one labeling them as the most frightening. 
We observed no statistically significant 
difference between the two, suggesting that participants are almost equally
frightened with a hacker stealing their medical records or their
sequenced genome. But, we did observe a statistically significant difference
between (D.1) and (D.4), suggesting that participants are not as frightened
by loss of genomic information as they are of identity theft (U $=230$, $n_1=n_2=16$, $P<0.001$, two-tailed).

\smallskip
We observed that several participants often used the degree of damage produced by
each incident involving information loss to identify the most frightening cases.
Experiment D also exhibited some interesting correlations with participants' demographics.
We observed that most users  with incomes lower than 50K
were most frightened by (D.3), i.e., their personal emails and pictures being stolen (5/6), as
opposed to very few users (1/10) with higher incomes ($\chi^2(1, N=32)=8.60 ~ p<0.01$). Whereas, users with incomes higher than 50K were most frightened by their identity being stolen (D.1), i.e., 
9/10 vs 1/6, resp., ($\chi^2(1, N=32)=8.60 ~ p<0.01$).

Next, we noticed that no privacy fundamentalist was
most frightened by loss of personal emails and pictures (0/5),
as opposed to a slight majority of pragmatists/unconcerned (6/11).
Whereas, privacy fundamentalists were most frightened by 
identity theft (5/5) as opposed to 5/11 for non-fundamentalists.
In both cases, it holds $\chi^2(1, N=32)=4.36 ~ p<0.05$, thus, there might be a statistically significant correlation between the Westin privacy index and being most frightened by either identity theft or loss of emails and/or pictures.

Reasons for such correlations can be explained by noticing that participants used the degree of damage
as the discriminant among information loss cases. Ostensibly, high-income users have more to lose (money-
and financial-wise) from identity theft -- e.g., P11 mentioned {\em ``It would cost me too much time and money.''}
Also, privacy fundamentalists might not be afraid of emails and pictures being stolen as they feel 
they take the appropriate precautions -- e.g., P2 reported {\em ``I always encrypt my sensitive emails and do not keep pictures on my laptop.''}

\subsection{Discussion}
In summary, our exploratory study unveils how complex is to evaluate the perception of security, privacy, and ethical issues related to the progress in Whole Genome Sequencing (WGS), and its medical and societal consequences.
We are aware that our analysis does not provide definite answers, as
we interviewed a relatively small number (16) of US-based volunteers, mostly high-educated and in the 25-34 age range.
Nonetheless, as suggested by~\cite{facio2011motivators} and~\cite{npr}, this seems to constitute the representative population of early adopters of personal genomic tests.

Therefore, while we expect our work to trigger further (and possibly larger) ethnographic studies in the field,
we do draw some preliminary conclusions: \vspace{0.1cm}

\begin{compactenum}
\item Labor and healthcare discrimination are top concerns among users, with an increase unfairness feeling associated with the former, and a lack of confidence in the protection granted by the law.
Fear of discrimination was predominant in our study, even more so than privacy concerns. %
\vspace{0.1cm} 
\item Users consistently raised the issue of control and preferred to retain and own data with their sequenced genome.
This suggests that design of genomics-related systems and applications should include mechanisms to let the users
feel in control of their data. \vspace{0.1cm}
\item Motivated by trust concerns, users prefer that doctors administered medical genetic tests, rather than specialized personal genomics companies. \vspace{0.1cm}
\item Perception of genetic tests is strongly related to the associated perceived medical benefit. Participants were mostly inclined to genetic tests that can help fight diseases, but less to tests that can help prevent them, and neutral w.r.t. discovering their ancestry. Although not totally unexpected, this indicates the need for identifying aligned incentives and for informed consent.\vspace{0.1cm}
\item Participants were more frightened by having their financial identity
and/or personal data stolen (with a bias toward the former for high-income, privacy fundamentalist users) than medical and genomic information. However, the perception of genomic information loss varied significantly among different participants and suggested that participants were more scared of their insurance provider or their employer using genomic information against them than of a hacker having access to it. \vspace{0.1cm}
\end{compactenum}

\section{Related Work}\label{sec:related}
The emergence of personal genomic tests and affordable Whole Genome Sequencing (WGS)  
motivate the need for better understanding related perception and concerns of involved users. However, much is left to be studied, since, to the best of our knowledge, there is no ethnographic study on Whole Genome Sequencing prior to our work.

Francke et al.~\cite{francke2013dealing} interviewed 63 customers of 23andMe.com  that tested for BRCA mutations. (BRCA gene mutations convey a high risk for breast and ovarian cancer). They analyzed customers' response among 32 mutation carriers (16 women, 16 men) and 31 non-carriers and concluded that direct access to BRCA mutation tests provided benefits to participants. None of the 25 users that had unexpectedly found out to carry the mutation reported extreme anxiety, and 4 experienced moderate, transitory anxiety, similar to the findings by Hamilton et al.~\cite{hamilton2009emotional}. 
Additional studies analyzing
patients' response to genetic tests for disease predisposition include
\cite{smith1995attitudes} (colon cancer), \cite{green2009disclosure} (Alzheimer's disease), \cite{bottorff2002women,bruno2004awareness} (breast cancer). %
Also, Andrews~\cite{andrews2001future} analyzed the concept of ``survivor's guilt'' experienced by individuals learning
not to carry a harmful mutation, while a family member does.

Brothers et al.~\cite{brothers2011two} found out that, 
when asked about an opt-out consent process, over 90 percent of participants agreed or strongly agreed that ``DNA biobank research is fine as long as people can choose not to have their DNA included.''
Some studies highlighted how privacy concerns are often an important obstacle to participation in large cohort studies~\cite{williams2009genetic}. Although 60 percent of people surveyed said they would participate in a study that involved storing data in biorepositories, 91 percent of those potential research participants would be concerned about privacy~\cite{kaufman2009public}. Another study
showed that, while participants trusted clinicians and researchers, they were concerned that results of genetic tests could end up in the wrong hands and be used against them~\cite{public-opinion}. 

Lapham et al.~\cite{lapham1996genetic} were among the first, in 1996, to analyze consumers' perspectives as to genetic discrimination and reported that people cited fear of losing insurance as a major reason to avoid genetic testing. However, discrimination by insurance companies was not a widespread reality in the 90s, as few of these cases had been filed
and even fewer had been won~\cite{hall2000patients}.
Many states have passed laws 
to protect medical (and also genetic) information, such as
the Health Insurance Portability and
Accountability Act (HIPAA), which provides a general framework for sharing and protecting Protected Health Information. 
In the U.S., there also exists legislation specific
against genetic discrimination -- the Genetic Information
Nondiscrimination Act (GINA) -- which prohibits discrimination on the basis of genetic information with respect to health insurance and employment \cite{wadman08}.
However, neither GINA or HIPAA placed any limits on health insurance rate setting.

Also, prior work analyzed
the issue of labor discrimination in relationship to disease predisposition, mostly from the legal standpoint. We refer to the work by Guttmacher et al.~\cite{guttmacher2003ethical} for details, along with a review of well-known rulings.

Ruiz et al.~\cite{ruiz2011research} compared the attitude of 279 patients from the United States and Spain who  had volunteered to donate a sample for genomic 
research, and showed that 48\% of participants would like to be informed about all individual results from future genomic studies using their donated tissue, especially those from the U.S. (71.4\%) and those 
believing that genetic information poses special risks 
(69.7\%). 
Trinidad et al.~\cite{trinidad2010genomic}
explored the attitude of research participants and possible future participants regarding Genome-Wide Association Studies  (GWAS): they found out that participants expressed a variety of opinions about the acceptability of wide sharing of genetic and phenotypic information for research purposes through large, publicly accessible data repositories. Most believed that making de-identified study data available to the research community was a social good to be pursued. Privacy and confidentiality concerns were common, although not necessarily precluding participation. Also, many participants voiced reservations about sharing data with for-profit organizations.

A recent report from the Presidential Commission for the Study of Bioethical Issues~\cite{president}
analyzed advances of Whole Genome Sequencing (WGS), and 
focused on potential privacy and ethical threats.
The report listed 12 recommendations
and provided a high-level effort to
identify and promote policies and practices that ensure scientific research, healthcare delivery, and technological innovation are conducted in a socially and ethically responsible manner.

Thus, we conclude that, while prior work has looked at responses to learning genetic test results
as well as discrimination concerns associated with genomics, 
no prior study has focused on the human factor in Whole Genome Sequencing.

\section{Conclusion}\label{sec:conclusion}
This paper presented the results of a series of semi-structured interviews involving 16 participants, that aimed to assess the perception of 
genetic tests, the attitude toward different Whole Genome Sequencing (WGS) programs, as well as their concerns with associated privacy and ethical threats.
The results of our interviews were analyzed both quantitatively and qualitatively and showed that
interviewees exhibited common fear of discrimination and demanded to retain strict control over their genetic information.
Our exploratory analysis suggests that the issues of control, trust, and discrimination are crucial aspects that the community should take into account,  not only from a policy standpoint, but also for user-centered designs of genomics-related applications that involve end-users (such as, ancestry and disease predisposition testing).

Our study highlights the need for more ethnographic studies in the field as well as the importance of informing the public with respect to privacy threats associated with genomic information disclosure and legal rights addressing discrimination issues. 

As part of future work, we plan to run follow-up user studies, relying on larger populations, that can further clarify some of the issues discussed in this paper. We are also working on strategies to raise awareness of privacy, ethical, and legal issues associated with WGS, as well as on user-centered designs of privacy-respecting genomic-testing platforms.

\descr{Acknowledgments.} The author would like to thank the participants who volunteered
for the study, Honglu Du for his valuable feedback,
and Darya Mohtashemi, Greg Norcie, and Julien Freudiger for their helpful comments.

\balance
\bibliographystyle{abbrv}
\bibliography{bibfile}

\end{document}